\documentclass[letterpaper, 10 pt, conference]{ieeeconf}
\IEEEoverridecommandlockouts                            
\usepackage{graphics} % for pdf, bitmapped graphics files
\usepackage{epsfig} % for postscript graphics files
\usepackage{times} % assumes new font selection scheme installed
\usepackage{amsmath} % assumes amsmath package installed
\usepackage{amssymb}  % assumes amsmath package installed
\usepackage[ruled,vlined]{algorithm2e}
\usepackage[utf8]{inputenc}
\usepackage[T1]{fontenc} 
\usepackage{cite}
\usepackage{amsmath,amssymb,amsfonts}
\usepackage{algorithmic}
\usepackage{graphicx}
\usepackage{textcomp}
\usepackage{xcolor}
\usepackage{comment}
\usepackage{tikz}
\usepackage{accents}
\usepackage{multirow}
\newtheorem{defi}{\textbf{Definition}}
\newtheorem{theorem}{Theorem}

\newtheorem{example}{Example}

\def\BibTeX{{\rm B\kern-.05em{\sc i\kern-.025em b}\kern-.08em
    T\kern-.1667em\lower.7ex\hbox{E}\kern-.125emX}}
\title{\LARGE \bf
Granger Causality for Predictability in Dynamic Mode Decomposition}
\author{G. Revati, Syed Shadab, K. Sonam, S. R. Wagh, and N. M. Singh%$^{1}$% <-this % stops a space
\thanks{G. Revati, Syed Shadab, S. R. Wagh, and N. M. Singh are with Control and Decision Research Centre (CDRC), EED,
        Veermata Jijabai Technological Institute, Mumbai 400019, India
        {\tt\small cdrc@ee.vjti.ac.in}. K. Sonam is with the Computer Science and Engineering Department, University of South Carolina, USA}%
 }
\begin{document}
\maketitle
\begin{abstract}
The dynamic mode decomposition (DMD) technique extracts the dominant modes characterizing the innate dynamical behavior of the system within the measurement data. For appropriate identification of dominant modes from the measurement data, the DMD algorithm necessitates ensuring the quality of the input measurement data sequences. On that account, for validating the usability of the dataset for the DMD algorithm, the paper proposed two conditions: Persistence of excitation (PE) and the Granger Causality Test (GCT). The virtual data sequences are designed with the hankel matrix representation such that the dimensions of the subspace spanning the essential system modes are increased with the addition of new state variables. The PE condition provides the lower bound for the trajectory length, and the GCT provides the order of the model. Satisfying the PE condition enables estimating an approximate linear model, but the predictability with the identified model is only assured with the temporal causation among data searched with GCT. The proposed methodology is validated with the application for coherency identification (CI) in a multi-machine power system (MMPS), an essential phenomenon in transient stability analysis. The significance of PE condition and GCT is demonstrated through various case studies implemented on 22 bus six generator system. 

\end{abstract}
\begin{keywords}
Coherency Identification, Dynamic Mode Decomposition (DMD), Granger causality, Hankel, Persistence of Excitation (PE).
\end {keywords}
\section{Introduction}
With the growing emphasis on data-driven modeling, understanding the interactions and connections among the time series drawn from observational data is a field of interest. Causality is the intersection of philosophy and sciences \cite{illari2011causality}, deriving the generalizations and theories from specific observations by analyzing the cause and effects among the observational data.
A primary approach for understanding the information flow amongst the time series is to determine the cross-correlation \cite{lutkepohl2005new} among the two time series and to discover the existence of a peak in the correlation at some non-zero lag.
The causal inferences drawn from the correlation are misleading since the correlation reveals only whether the two variables are statistically linked.
The causal relationship amongst two variables can be direct, or indirect due to confounding effect \cite{peters2017elements} i.e., besides the variables under study there is some additional unnoticed variable correlated with the considered variables. Furthermore, the correlation being a symmetric measure fails to provide any information about the causality direction.

 As per the principle of time asymmetry of causation, classical physics employs the precedence of causes over effects. 
  Accounting for the direction of causation, a dynamical model identification is another approach.
 The concept of dynamical model identification is fundamentally developed on the fact that law drives the system and enables the evolution of the same state in a similar manner \cite{tissot2014granger} i.e., similar effects are produced by the same causes mentioned as per physical determination. 
  The laws defining the system dynamics are identified from the regression of observational data achieved through the evaluation of correlation.
  For detecting and quantifying the temporal causality amidst the time series, a powerful statistical test known as Granger Causality (GC) was first proposed in \cite{granger1969investigating}.
   The widespread applications of the GC in neuroscience \cite{roebroeck2005mapping}, economy \cite{comincioli1996stock}, and climate modelling \cite{elsner2007granger} are mentioned in the literature.
 The fundamental notion behind GC is the enhancement in the prediction of one variable with the introduction of past information of another variable along with the past information of the considered variable itself.

Conventionally the control applications extensively opted for the system identification methods fitting the data to the model parameterized priori \cite{ljung1998system}. The growing complexity and huge amount of available system data challenged the standard strategies for learning the dynamical system. Alternatively, the paradigm shift occurred towards the identification of a dynamical system from the raw measurement data of the system. The data-driven modeling approaches are generally dependent on searching for the accurate combination of the known trajectory in order to achieve a reliable prediction which is usually an ill-conditioned problem \cite{yin2021maximum}. For dealing with such problems, the Moore-Penrose pseudoinverse \cite{sedghizadeh2018data} solving the least norm problems is preferred due to computational simplicity. One such data-driven subspace prediction strategy is the Dynamic Mode Decomposition (DMD) \cite{kutz2016dynamic} which decomposes the high dimensional data into spatiotemporal coherent modes.  

DMD is a dimensionality reduction technique \cite{9654641} pioneered in the fluid dynamics community by Peter Schmid for identifying the linear approximation from the data comprising the dominant modes describing the dynamical behavior of the system \cite{arbabi2017ergodic}. The quality identification of the dominant modes capturing the dynamics depends on the quality of the measurement data exploited for the strategy. For capturing the modes of the system, the cardinality of the measurement sequences utilized in the DMD should be greater than or equal to the underlying system modes. Hence the dimensions of the subspace spanning the essential dynamical modes are increased with the Hankel matrix \cite{9480170} introducing the new state variables. The temporal evolution of the system lies in the column space and the row space defines the spatial structure of the system modes hence for the estimation of an accurate approximate linear model a sufficient number of rows and columns of the input data matrices are necessary. 
Hence for identifying the accurate linear model with predictability, the paper has proposed two conditions: Persistence of excitation (PE) and Granger Causality Test (GCT). The PE is a necessary condition and provides the lower bound on the trajectory length and the GCT which being a sufficient condition is informative to detect the causation among the measurement sequences and to find the appropriate order of the model ensuring the predictability of the identified linear model.

The suitability of the proposed approach is verified with the application of coherency identification in a multimachine power system (MMPS).  Coherency is a property of generators to swing together the coherency identification is a necessary phenomenon for the transient stability analysis in MMPS. The relevance of PE condition for capturing the dominant dynamical modes of the system and the significance of GCT to establish the vital role of causation to ensure predictability is demonstrated with the various experimental case studies implemented on the 22 bus 6 generator power system. 

The remaining paper is structured as follows: The concepts including Dynamic mode decomposition, the persistence of excitation, vector autoregression, and Granger causality are discussed briefly as preliminaries in Section \ref{Preliminary}. The proposed methodology explaining the details of the PE condition and the Granger Causality Test
along with the DMD algorithm is presented in Section \ref{Proposed Method}. The results for the comprehensive case study along with the test and error analysis are illustrated in Section \ref{results}. The paper is concluded with the future work in Section \ref{conclusion}.

\section{Preliminaries}
\label{Preliminary}
\subsection{Dynamic Mode Decomposition (DMD)}
DMD is designed to extract spatially coherent modes, oscillate at a fixed frequency, and decay or growth at a fixed rate \cite{katayama2005subspace}.
DMD is a data-driven technique that emerged from the fluid dynamics community \cite{schmid2010dynamic} identifying a dynamical system from the observational data. 
DMD is strongly related to the Koopman operator theory (KOT) which provides an infinite dimensional linear representation $\mathcal{K}$ of the nonlinear system dynamics $\varphi$ acting on the finite-dimensional manifold $\mathcal{M}$.
DMD is an approach that seeks the $\mathcal{A}$ matrix such that its spectrum approximates the spectrum of the Koopman operator. The dominant eigenvalues and eigenvectors of this $\mathcal{A}$ matrix are very informative \cite{9597702} about the dynamical attributes of the system such as the frequency, decay, growth, and flow modes.

Consider a set of pairs of observations $\mathrm{x_j}$ and $\mathrm{y_j}$, $\mathrm{j}=1, 2,\cdots, \mathrm{n}$ related temporally such that
\begin{equation}
    \mathrm{y_j(z)}=\mathrm{x_j}(\varphi(\mathrm{z}))
    \label{dynamical system}
\end{equation}
where $\mathrm{z}\in \mathcal{M}$, $\varphi :\mathcal{M}\rightarrow \mathcal{M}$ is a dynamical system, and $\mathrm{x_j,y_j}: \mathcal{M}\rightarrow \mathbb{R}$. $\mathrm{y_j}$ is one step ahead in time that of $\mathrm{x_j}$ i.e. if $\mathrm{x_j}$ denotes the observation at time $\mathrm{t}$, then $\mathrm{y_j}$ represents the measurement at time $\mathrm{t+\Delta t}$. From these measurement time series, two data matrices $X$ and $Y$ are constructed as 
\begin{equation}
    \mathrm{X}=\begin{bmatrix}
\mathrm{x_1} &\mathrm{x_2}  & \cdots   & \mathrm{x_n}
\end{bmatrix},
\hspace{0.6 cm}
\mathrm{Y}=\begin{bmatrix}
\mathrm{y_1} &\mathrm{y_2}  & \cdots   & \mathrm{y_n}
\end{bmatrix}
\label{input matrices}
\end{equation}
also $\mathrm{X},\mathrm{Y} \in \mathbb{R}^{\mathrm{m}\times \mathrm{n}}$, where $\mathrm{m}$ is the dimension of the manifold $\mathcal{M}$. The objective of the DMD algorithm is to evaluate the approximation $\mathcal{A}$ such that 
\begin{equation}
  \mathrm{Y}=\mathcal{A}\mathrm{X} 
  \label{DMD aim}
\end{equation}
The analytical solution of the above problem is given by 
\begin{equation}
    \mathcal{A}=\mathrm{Y}\mathrm{X}^\dagger 
    \label{pseudoinverse}
\end{equation}
 As the input matrices involved in the computation of $\mathcal{A}$ matrix are rectangular the Moore-Penrose pseudo-inverse $\mathrm{X}^{\dagger}$ is used.
The solution \ref{pseudoinverse} originates from the least square problem of minimizing the error
\begin{equation}
    \left \| \mathrm{Y}-\mathcal{A}\mathrm{X} \right \|_{F}
\end{equation}
$\left \| .\right \|_{F}$ is a Frobenius norm given as
\begin{equation}
    \left \| Z\right \|_{F}=\sqrt{\sum_{j=1}^{p}\sum_{k=1}^{q}Z_{jk}^2}
\end{equation}

Practically the computation of $\mathcal{A}$ is challenging due to the very large size of observations i.e.  $\mathrm{m>n}$ resulting in an under-determined system. To alleviate this difficulty, the spatial dimensions of the input data are reduced through Proper Orthogonal Decomposition (POD). The linear subspace spanned by the set of $r$ orthogonal modes approximates the space $\mathbb{R}^\mathrm{m}$ sufficiently to achieve the dimensionality reduction.
The proper orthogonal modes are evaluated via the Singular Value Decomposition (SVD) of $\mathrm{X}$ given as 
\begin{equation}
    \mathrm{X}=\mathrm{U}\mathrm{S}\mathrm{V}^*
    \label{SVD}
\end{equation}
with $\mathrm{U}\in\mathbb{R}^{\mathrm{m}\times r} $ whose each column represents the eigenvector of $XX^T$, $\mathrm{S}\in\mathbb{R}^{r\times r}$ is the diagonal matrix having eigenvalues of $X$ in descending order and $\mathrm{V}\in\mathbb{R}^{\mathrm{n}\times r}$ is the matrix in which each row represents the eigenvector of $XX^T$. The $r$ proper orthogonal modes correspond to the dominant left singular vectors ($\mathrm{U}_r$) of SVD associated with the dominant singular values.  

The SVD (\ref{SVD}) aids to search for the reduced order subspace containing the dominant system modes through the projection of $\mathcal{A}$ onto the POD modes.
$\tilde{\mathcal{A}}=\mathrm{U}^*\mathcal{A}\mathrm{U}=\mathrm{U}^*\mathrm{Y}\mathrm{V}\mathrm{S}^{-1}$. The DMD modes and eigenvalues are evaluated from the eigendecomposition of $\tilde{\mathcal{A}}$, i.e., $\tilde{\mathcal{A}}\mathrm{W}=\mathrm{W}\Lambda$. The exact DMD modes $\Phi $ are evaluated by transforming back to the original space of higher dimensions, i.e. $\Phi=\mathrm{Y}\mathrm{V}\mathrm{S}^{-1}\mathrm{W}$. The temporal evolution of the modes is identified with the eigenvalues from the diagonal of matrix $\Lambda$.

% DMD assumes multichannel time series being generated by autonomous time-invariant systems where each channel is a state variable.  If these state variables do not span the time series principal modes the DMD algorithm fails \cite{8718360}. Arranging the signal in block Hankel matrix before performing DMD, the order of the autonomous system is increased enabling the DMD to capture more signal modes.  

\subsection{Persistence of excitation (PE)}
For the linear time-invariant system which is controllable, if a component of a state is PE of an adequately higher order, then the sections of the trajectory span the space characterizing the behavior of the system \cite{willems2005note}.

In order to ensure the consistency of the model estimated during the identification experiments \cite{shadab2023finite} \cite{shadab2022gaussian}, the data sequences used for the subspace identification must be PE of significant order. For efficiently identifying the system modes from the deterministic input data sequences \cite{5990975}, the virtual data sequences analogous to the multi-variable system fulfilling the PE condition are designed with the hankel block matrix representation of the input data sequence.

Let $x\in \mathbb{R}^{n}$ be a state vector denoted as $x_{[k,k+S]}$, where $k\in X $ is the time instant for a first sample and $S\in N$ are the total samples taken. The state vector in the interval $[k,k+S]\cap X$ is defined as
 
\begin{equation}
    x_{[k,k+S]}=\begin{bmatrix}
\\ x(k)
\\ x(k+1)
\\ \vdots 
\\x(k+S)

\end{bmatrix}
\end{equation}
The state vector $x_{[k,k+S]}$ organized in the Hankel matrix is represented as
\begin{equation}
    \begin{bmatrix}
x(k) & x(k+1) & \cdots  &x(k+S-M+1) \\ 
 x(k+1)&x(k+2)  & \cdots  &x(k+S-M+2) \\ 
\vdots  & \vdots  & \ddots  & \vdots \\ 
 x(k+M-1)& x(k+l) & \cdots  & x(k+S)
\end{bmatrix}
\label{hankel}
\end{equation}
where $k\in X,S,M\in N$. The total number of system eigenvalues determines the value of $M$

\begin{defi}
A signal trajectory $ x_{[k,k+S]}$  is PE of the order $L$, if the block hankel matrix in (\ref{hankel}) has a full rank $nL$
\label{PE}
\end{defi}

Basically, the definition indicates that to satisfy the PE condition, the section of the trajectory of length $L$ must be adequately long enough to excite the controllable system modes in the window of $L$ and reproduce them. The Definition \ref{PE} gives the lower bound on the trajectory length such that $m>= (n+1)L-1$, the number of rows will be lesser or equal to the columns i.e. there will be more temporal samples than the spatial samples \cite{9143708}.

\subsection{Vector Auto regression (VAR)}
A multivariate time series $x_1, x_2, \cdots, x_m$, with each measurement being a n-dimensional vector consisting the elements $x_{1t}, x_{2t}, \cdots, x_{nt}$ is realised with a vector stochastic process $X_1, X_2, \cdots$. 
This time series can be modeled with a Vector Autoregressive (VAR) of order $p$ represented as
\begin{equation}
X_t=\sum_{k=1}^{p} A_k X_{t-k} + \epsilon_{t}
\label{VAR def}
\end{equation}
$A_k\in \mathbb{R}^{n\times n}$ is a regression coefficient matrix and $\epsilon_t$ is the white noise corresponding to the residuals. 
VAR models detect the simultaneous evolution patterns of multivariate time series.
The objective of the VAR model is to find the coefficient matrix demonstrating the temporal correlation between the multivariate time series. The predictive VAR model represents the value of $x_t$ at any time $t$ as a combination of its past values. The predictable patterns in the data are captured with the coefficient matrix whereas the unpredictable part is accounted for with the residual terms.

GC analysis draws causal inferences among various variables based on their VAR representation. To obtain a valid analysis of GC, the coefficients of the VAR model (\ref{VAR def}) should be square summable and stable \cite{hamilton2020time} \cite{lutkepohl2005new}. Square summability implies that $\sum_{k=1}^{p}\left \| A_k \right \|^2<\infty$ i.e.  even for the infinite order of the model, the regression coefficients are not blowing up. The stability is related to the characteristic polynomial of coefficient matrix $A_k$, $\varphi (z)=\left | I-\sum \sum_{k=1}^{p}A_kz^k \right |$ where $Z\in\mathbb{C}$. If the characteristic polynomial of the coefficient matrix is invertible on the unit disc $\left | z \right |\leq 1$ in the complex plane \cite{lutkepohl2005new}, then the coefficients of the VAR model are stable.

\subsection{Granger Causality (GC)}

Granger causality (GC) is a framework established on the temporal precedence implying that causes precede their effects. Temporal precedence being the innate characteristic of the time series, the GC helps to establish the causal relationship among the time series inferring that the past is causing the future. The physical interpretation of the causality is that the causes are responsible for the unique changes in the corresponding effects, i.e. the causal series contains the unique information about the effect series which is not available otherwise \cite{granger1969investigating}. In fact, GC does not actually indicate the true causality, rather it examines the influence of one series on the forecasting ability of another series.

Considering two time series $x_t$ and $y_t$ the GC finds the causal relationship amidst $x_t$ and $y_t$ depending on whether the past values of $x_t$ assists in the prediction of $y_t$ conditional on having already considered the effect of the past values of $y$ on the prediction of $y_t$. 

\begin{defi}
Suppose $\mathrm{H}_{<t}$ be representing the history of all relevant information available up to $(t-1)$, $\mathrm{Pr}(x_t\mid \mathrm{H}_{<t})$ be denoting the prediction of $x_t$ given $\mathrm{H}_{<t}$, the GC suggests $y$ to be causal for $x$ if

\begin{equation}
\mathtt{var}\left [ x_t-\mathrm{Pr}(x_t\mid \mathrm{H}_{<t}) \right ]<
\mathtt{var}\left [ x_t-\mathrm{Pr}(x_t\mid \mathrm{H}_{<t}\setminus y_{<t}) \right ]
\label{GC}
\end{equation}
 whereas $\mathrm{H}_{<t}\setminus y_{<t}$ specifies that the values of $y_{<t}$ are excluded from $\mathrm{H}_{<t}$. Equation (\ref{GC}) describes that the variance of the prediction error of $x$ with the inclusion of the history of $y$ is reduced \cite{shojaie2022granger} consequently inferring that the past values of $y$ enhance the prediction of $x$.
 \label{GC_defi}
\end{defi}

The primary argument of GC is based on the identification of a unique linear model from the data. Let $x$ and $y$ be the two time series of length $N$, and described by the uni-variate vector auto-regressive (VAR) model of order $p<N$, given as
\begin{equation}
    x(t_k)=\sum_{j=1}^{p}\alpha _jx(t_{k-j})+\eta (t_k)
    \label{UVAR}
\end{equation}
The time series $x$ is also represented with the bi-variate VAR model, given as
\begin{equation}
    x(t_k)=\sum_{j=1}^{p}\tilde{\alpha} _jx(t_{k-j})+\sum_{j=1}^{p}\tilde{\beta} _jy(t_{k-j})+\tilde{\eta} (t_k)
    \label{BVAR}
\end{equation}
where $\alpha$, $\tilde{\alpha}$ and $\tilde{\beta}$ be the model coefficients, and $\eta$, $\tilde{\eta}$ are prediction errors corresponding to each VAR models. According to the definition of GC, if the prediction error of the bivariate VAR model is less than the prediction error of the univariate VAR model, then it implies that $y$ "granger causes" $x$.

For quantifying GC the Granger Causality Index (GCI) is introduced in \cite{heyse2021evaluation} as the logarithmic ratio of the prediction errors of the VAR models (\ref{UVAR}) and (\ref{BVAR}). Mathematically, GCI is given as 
\begin{equation}
    \mathcal{F}_{2\rightarrow 1}=ln\left ( \frac{var(\eta)}{var(\tilde{\eta})} \right )
\end{equation}
 If the inclusion of past values of $y$ does not improve the prediction of $x$, then $var(\eta)\approx var(\tilde{\eta})$, and hence $\mathcal{F}_{2\rightarrow 1}\approx0$. With the enhanced prediction, the variance of the bivariate VAR model reduces resulting in GCI greater than zero. Larger values of GCI indicate a stronger causal relationship. 
%\subsection{Granger Causality Test (GCT)}
%\section{Diagonal Averaging with Hankel Blocks}

\section{Granger Causality for predictability in Dynamic Mode Decomposition}
\label{Proposed Method}
\begin{figure*}[ht!]
    \centering
    \includegraphics[width=\linewidth]{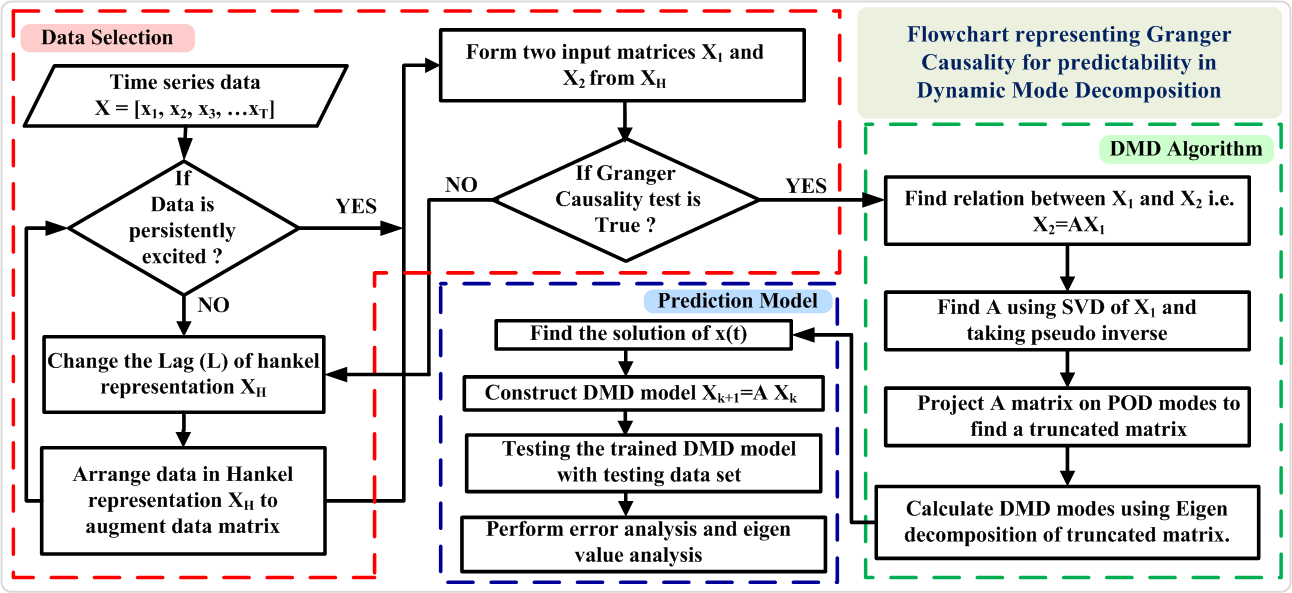}
    \caption{Flowchart demonstrating Granger Causality for predictability in Dynamic Mode Decomposition}
    \label{Flowchart}
\end{figure*}

The objective of DMD is to evaluate the spatiotemporal coherent modes from the measurement data matrices. The temporal modes (reflected from the dominant eigenvalues) lie in the column space and the spatial modes (reflected from the dominant eigenvectors) lie in the row space. With the hankel representation, new state variables are created by increasing the number of rows, but this results in a reduction of the number of columns hampering the identification of temporal modes. Similarly, if the number of columns in the hankel matrix is increased then the total number of measurements being constant the number of rows is reduced resulting in the reduction of state variables and hence reduction in the model order hampering the identification of system modes. Hence for dominant modes detection and estimation of an accurate approximate linear model, the number of rows and columns of the data matrix should be adequate.

The PE condition provides the lower bound on the trajectory length assuring the minimum number of columns required for capturing the temporal modes. And the Granger Causality Test (GCT) offers the order of model which will give the minimum prediction error. The order of the model in turn gives the information about the number of new state variables required to be introduced so as to identify the system modes. Hence only satisfying the PE condition is not enough to guarantee the estimation of an accurate linear model having the capability of predictability. The accuracy and predictability are ensured with the GCT. On that account, PE becomes the necessary condition, and GCT becomes the sufficient condition for the Frobenius norm optimization problem in order to guarantee the estimation of an accurate linear model with predictability.

Usually, the time series data is not persistently excited and is unsuitable for the identification of appropriate dominant spatiotemporal coherent modes characterizing the innate system dynamics. Therefore, before implementing the DMD algorithm on the data, it is preliminary to select the proper data set. The proposed methodology is hence divided into the following steps data selection, DMD algorithm, and prediction model analysis as demonstrated in Fig \ref{Flowchart}.

\subsection{Data Selection}

To validate the usability of the dataset for the DMD algorithm, it must satisfy two conditions; one is necessary and the other is sufficient. The data should be sufficiently rich to capture the system dynamics which is checked by the PE condition which is a necessary condition. But it does not guarantee the predictability of the identified model. For the dataset to ensure predictability, it should qualify for the GCT.

\subsubsection{Necessary Condition: Persistence of excitation (PE)}
For extracting the essential modes of the system capable of predicting the future behavior of the system the cardinality of the set of linearly independent signals comprising the system modes should be greater than or equal to the number of dominant modes. The temporal evolution of the system is captured in terms of the column vectors therefore the primary objective is to present these column vectors as a linear combination \cite{tirunagari2017dynamic} providing the prediction of system dynamics. 

DMD is fabricated to work with the low-rank structures but the linear dependencies present in the measurement data prevent spanning the subspace containing the dominant modes. The issue is resolved by artificially expanding the dimensions of subspace spanning the essential system modes by adding the time-lagged samples through the Hankel matrix \cite{8718360}.   
The block hankel matrix redesigns the input measurement sequences into a virtual data sequence incorporating new state variables enabling the approximate linear model identification with the low-rank structure.

The quality of the linear approximation achieved through the DMD algorithm depends on the data quality such that the reduced order models in DMD extracted from the enriched data must illustrate the true dynamics of the system.
% Let the measurement vector be 
% \begin{equation}
%     X=\begin{bmatrix}
% x(1) & x(2) & \cdots & x(m)
% \end{bmatrix}
% \end{equation}
% where $x\in \mathbb{R}^n$ is the measurement sample recorded at the time instant $t_k$, $k=1, 2, \cdots,m$.

% To enrich the measurement data sequence the hankel matrix of the PE order $L$ is formulated such that
% \begin{equation}
%     X_H=\begin{bmatrix}
% x(1) & x(2) & \cdots & x(m-L+1)\\ 
% x(2) & x(3) & \cdots  & x(m-L+2)\\ 
%  \vdots & \vdots  & \ddots  &\vdots  \\ 
% x(L) &x(L+1)  & \cdots  & x(m)
% \end{bmatrix}
% \end{equation}
%  To proceed with DMD two input matrices $X_1$ and $X_2$ are formulated from the matrix $X_H$.
%  \begin{equation}
%      X_1=\begin{bmatrix}
% x(1) & x(2) & \cdots & x(m-L)\\ 
% x(2) & x(3) & \cdots  & x(m-L+1)\\ 
%  \vdots & \vdots  & \ddots  &\vdots  \\ 
% x(L) &x(L+1)  & \cdots  & x(m-1)
% \end{bmatrix}
%  \end{equation}
% \begin{equation}
%     X_2=\begin{bmatrix}
% x(2) & x(3) & \cdots & x(m-L+1)\\ 
% x(3) & x(4) & \cdots  & x(m-L+2)\\ 
%  \vdots & \vdots  & \ddots  &\vdots  \\ 
% x(L+1) &x(L+2)  & \cdots  & x(m)
% \end{bmatrix}
% \end{equation}
Let $x_{k+1}=Ax_k$ be a discrete-time, linear dynamical system generating the time series of data  
\begin{equation}
X=\begin{bmatrix}
x(1) & x(2) & \cdots & x(m)
\end{bmatrix}
\end{equation}

 where $x\in \mathbb{R}^n$ is the measurement sample recorded at the time instant $t_k$, $k=1, 2, \cdots,m$.

The measurement data sequence is enriched with the help of the hankel matrix of order $L$ such that
\begin{equation}
    X_H=\begin{bmatrix}
x(1) & x(2) & \cdots & x(m-L+1)\\ 
x(2) & x(3) & \cdots  & x(m-L+2)\\ 
 \vdots & \vdots  & \ddots  &\vdots  \\ 
x(L) &x(L+1)  & \cdots  & x(m)
\end{bmatrix}
\end{equation}
For simplicity, the data matrices $X_1$ and $X_2$ are formulated from $X_H$ as
 \begin{equation}
     X_1=\begin{bmatrix}
x(1) & x(2) & \cdots & x(m-L)\\ 
x(2) & x(3) & \cdots  & x(m-L+1)\\ 
 \vdots & \vdots  & \ddots  &\vdots  \\ 
x(L) &x(L+1)  & \cdots  & x(m-1)
\end{bmatrix}
 \end{equation}
\begin{equation}
    X_2=\begin{bmatrix}
x(2) & x(3) & \cdots & x(m-L+1)\\ 
x(3) & x(4) & \cdots  & x(m-L+2)\\ 
 \vdots & \vdots  & \ddots  &\vdots  \\ 
x(L+1) &x(L+2)  & \cdots  & x(m)
\end{bmatrix}
\end{equation}

The exact low-rank variant of the above-mentioned dynamical system is computed from the SVD of the original state transition matrix (STM) (i.e. $A=U\Sigma V^*$. The exact low rank STM is computed by truncating to the $r$ significant singular values in $\Sigma$ and the respective modes from $U$ and $V$ (i.e. $A_r=U_r \Sigma_r V_r^*$. The respective low rank dynamical system is $x^r_{k+1}=A_rx_k^r$ and the low rank data matrices are denoted as $X_{1}^r$ and $X_{2}^r$. 
By definition
\begin{equation}
    \left \| X_2-AX_1 \right \|=0
\end{equation}
and
\begin{equation}\label{truncated STM}
    \left \| X_{2}^r-A_rX_{1}^r \right \|=0
\end{equation}
To prove the PE condition, another operator $\bar{A}$ i.e. DMD approximated low-rank STM is defined. With SVD and Moore-Penrose pseudo-inverse, the $\bar{A}$ matrix of a particular rank is evaluated such that the Frobenius norm $\left \| X_2-\bar{A}X_1 \right \|$ is minimized \cite{golub1987generalization} as assured by the Young-Eckart Theorem. 
\begin{theorem}
If the state data sequence $X$ fulfills the PE condition, then the DMD estimated STM $\bar{A}$ is the best possible low-rank estimate $A_r$ of the exact STM $A$.
\end{theorem}
\textbf{Proof:}
If the signal $X^r$ originated from the dynamical system $x^r_{k+1}=A_rx_k^r$ is PE, then an appropriate choice of gradient based law will assure that $\left \| \bar{A}-A_r \right \|\rightarrow 0$ exponentially fast.
This implies that 
\begin{equation}\label{intermediate}
    \left \| \bar{A}X_1^r-A_r X_1^r\right \|\rightarrow 0
\end{equation}
as $X_1^r$ is constant.
Inclusion of $(X_2^r-X_2^r)=0$ in (\ref{intermediate}) has no effect, hence
\begin{equation}
    \left \| (X_2^r-\bar{A}X_1^r)-(X_2^r-A_rX_1^r) \right \|\rightarrow 0
\end{equation}
From (\ref{truncated STM}) $\left \| X_{2}^r-A_rX_{1}^r \right \|=0$,
so
\begin{equation}
    \left \| (X_2^r-\bar{A}X_1^r) \right \|\rightarrow 0
\end{equation}

Hence, the convergence of DMD to the best low-rank approximation of STM of the full-state system is implied through the PE condition.
 When PE is satisfied the resulting approximate lower order model extracted with DMD becomes more robust to the inputs since the model almost depicts the actual dynamical behavior of the system. Hence the order of PE $L$ is specified such that with the enriched data the quality of $\bar{A}$ will improve reducing the Frobenius norm nearly to zero.

\subsubsection{Sufficient Condition: Granger Causality Test (GCT)}
Since the PE condition does not guarantee that the Frobenius norm $\left \| X_2-\bar{A}X_1 \right \|$ is identically zero, the PE condition alone is not sufficient. Hence the GCT is proposed to identify the causal relationships among the data to improve the predictability of the linear model estimated via DMD.

 Consider a time series $x(t)=(x_1(t),x_2(t),\cdots,x_n(t))\in \mathbb{R}^n$ being represented by a VAR model of order p 
 \begin{equation}
     x(t)=c+ A_1 x(t-1)+ A_2 x(t-2)+ \cdots+ A_p x(t-p) + \epsilon(t)
     \label{vartimeseries}
 \end{equation}
with $c\in \mathbb{R}^n$ be a constant vector, $(A_1, A_2, \cdots, A_p) \in \mathbb{R}^{n \times n}$ be regression coefficients,  and $\epsilon$ be a gaussian noise process with variance $\Sigma$.
With reference to the considered VAR model (\ref{vartimeseries}), the GC of the model is written in terms of a linear equation as, i.e., if $x_j$ does not granger cause $x_i$ then
\begin{equation}
    (A_{k})_{ij}= 0, \hspace{0.5 cm} k=1, 2, \cdots, p 
\end{equation} 
The underlying GC structure among the series is identified from the zero patterns from the estimated VAR coefficient matrix \cite{raksasri2017guaranteed}. 

\begin{example}

Consider a time series $x(t)\in \mathbb{R}^3$, where the causal relations between the elements of the series are represented in Fig. \ref{GC VAR}.
\begin{figure}[ht!]
    \centering
    \includegraphics[width=0.8\linewidth]{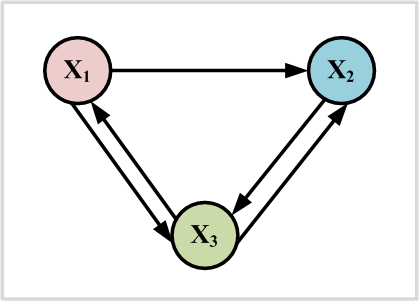}
    \caption{Causal relationship between the elements of series}
    \label{GC VAR}
\end{figure}
From the Fig. \ref{GC VAR}, it is observed that $X_1$ Granger causes $X_2$ and $X_3$, $X_3$ Granger causes $X_1$ and $X_2$, $X_2$ Granger causes $X_3$, and $X_2$ does not Grager cause $X_1$. Hence the given time series is represented with a VAR (4) model of order $p=4$ as follows
\begin{align}\label{GC_VAR_example}
    \begin{split}
        \begin{bmatrix}
x_1(t)\\ 
x_2(t)\\ 
x_3(t)
\end{bmatrix}=c+&\begin{bmatrix}
\times  & 0  &\times \\ 
 \times & \times &  \times\\ 
 \times & \times  & \times 
\end{bmatrix} 
\begin{bmatrix}
x_1(t-1)\\ 
x_2(t-1)\\ 
x_3(t-1)
\end{bmatrix}+\begin{bmatrix}
\times  & 0  &\times \\ 
 \times & \times &  \times\\ 
 \times & \times  & \times 
\end{bmatrix} \\& \begin{bmatrix}
x_1(t-2)\\ 
x_2(t-2)\\ 
x_3(t-2)
\end{bmatrix}+\begin{bmatrix}
\times  & 0  &\times \\ 
 \times & \times &  \times\\ 
 \times & \times  & \times 
\end{bmatrix} \begin{bmatrix}
x_1(t-3)\\ 
x_2(t-3)\\ 
x_3(t-3)
\end{bmatrix}+\\&\begin{bmatrix}
\times  & 0  &\times \\ 
 \times & \times &  \times\\ 
 \times & \times  & \times 
\end{bmatrix}  \begin{bmatrix}
x_1(t-4)\\ 
x_2(t-4)\\ 
x_3(t-4)
\end{bmatrix}+\epsilon(t)
    \end{split}
\end{align}
\end{example}

In above equation (\ref{GC_VAR_example}), ($\times$) represents the non zero entries in the coefficient matrix. The zero entry in the coefficient matrix represents the lack of causal relationship and the variable $X_1$ is not granger causing $X_2$, whereas all the non-zero entries ($\times$) in the coefficient matrix represent the causal relationship among the respective variables.

 While estimating the VAR model required for GC analysis, the number of time lags to be included, i.e., the order of the model, is an important parameter to select. Very low model order leads to the poor representation of the measurement data leading to the failure of capturing the adequate dominant system modes while a very high order of the model leads to an over-fitted model with high prediction errors \cite{lutkepohl2005new}. Hence the model order should be selected by accounting for the trade-off between the good level of representation of the measurement data and the low order representation. The extensive approach for the appropriate model order selection is to minimize the criterion \cite{seth2010matlab} which balances the variance accounted for by the model, against the number of coefficients to be estimated. Model selection is achieved through two most popular information criteria: the Akaike Information Criterion (AIC) \cite{akaike1974new} and the Bayesian Information Criterion (BIC) \cite{schwartz1978estimating}.
For $n$ variables, with the $\Sigma$ as covariance matrix the model order $p<T$ is evaluated as
\begin{equation}
    AIC(p)=ln(det(\Sigma))+\frac{2pn^2}{T}
\end{equation}
\begin{equation}
    BIC(p)=ln(det(\Sigma))+\frac{ln(T)pn^2}{T}
\end{equation}

With the VAR model of appropriate order, the GCT is performed to detect the temporal causality. 
GCT is a statistical test developed to quantify the temporal causal effect among the time series.
Hypothesis testing, particularly the Wald test, is formulated to detect the GC among various time series based on the fundamental idea of capturing the zero value structure existing in the coefficient matrix of the VAR model of the series (i.e. if a particular $(i,j)^{th}$ element of matrix $A_k$ is zero then it implies that from the GC perspective no relation exists between the $i^{th}$ and $j^{th}$ elements of the series). Wald test is characterized by the restriction function explaining the hypothesis and the Wald statistic which is a quadratic form of the restriction function. The intuition behind the test is that if the actual value of the parameters is zero, then their estimated value should be significantly very small.

Wald test is designed to assess the GC by testing if the hypothesis $(\hat{A_k})_{ij}=0$ is true. 
Every entry of the coefficient matrix $A$ is tested with the restriction function given as
\begin{equation}
    \beta(\hat{\theta })=\left ( (\hat{A_1})_{ij}, (\hat{A_2})_{ij}, \cdots, (\hat{A_p})_{ij} \right )
\end{equation}
whereas $\theta$ is a vectorization of the coefficient matrix $A$ and $\hat{A}$ is the estimation of the coefficients. The null hypothesis for the Wald test is proposed as
\begin{equation}
    H_0:\beta(\hat{\theta})=0
\end{equation}
The Wald test statistics involving the quadratic terms of the restriction function are defined as 
\begin{equation}
    W_{ij}=\beta (\hat{\theta })^T\left [ \widehat{Avar}(\hat{\theta })_{ij} \right ]^{-1}\beta(\hat{\theta })
\end{equation}
where $\widehat{Avar}(\hat{\theta })_{ij}$ represents the $p\times p$ asymptotic covariance matrix of the estimate of parameter $\theta$. Given the $p$ degrees of freedom, under the null hypothesis $H_0$ the Wald test distribution converges to the Chi-square distribution.

Considering the critical value $\chi _\alpha^2$ the significance level is evaluated as the probability that the test statistics is greater than the critical value i.e. $W_{ij}>\chi _\alpha^2$. 
For every element $(i,j)$ of the coefficient matrix, the Wald statistics are calculated and compared with the significance level. 
When the test statistics exceed the significance level, the null hypothesis $H_0$ is rejected, avoiding the possibility of noncausality. Consequently rejecting the null hypothesis implies that the corresponding element in the coefficient matrix is non zero indicating the presence of granger causality among the variables. With the Wald statistics values corresponding to each element $A_{ij}$ compiled together in a matrix a Wald statistic matrix is formulated which later on compared with the significance value provides the binary matrix of test results containing only 1 or 0 entry. If the $(i,j)^{th}$ entry of the result matrix is one then it implies that the component $x_j$ granger causes $x_i$ whereas zero indicates that the component $x_j$ does not granger cause $x_i$.

\subsection{DMD Algorithm} 
 with the enriched data set satisfying both necessary (PE) and sufficient (GCT) conditions, the approximate linear model with appropriate order is estimated further through the DMD algorithm. For the evaluation of Frobenius norm $\left \| X_2-AX_1 \right \|$ the pseudo-inverse of $X_1$ matrix is required which is achieved through the Singular Value Decomposition (SVD) of the matrix $X_1$.
 \begin{equation}
     X_1=U\Sigma  V^*
     \label{X1SVD}
 \end{equation}
where $U\in \mathbb{C}^{n\times r}$, $\Sigma\in \mathbb{C}^{r\times r}$, and $V\in \mathbb{C}^{m\times r}$. 
The reduced order model is extracted by projecting the state transition matrix $A$ onto the Proper Orthogonal Decomposition (POD) modes evaluated by the SVD (\ref{X1SVD}).
\begin{equation}
    \tilde{A}=U^*AU=U^*X_2V\Sigma ^{-1}
\end{equation}
The reduced rank state transition matrix $\tilde{A}$ represents the best fit matrix minimizing the Frobenius error norm. $\tilde{A}$ is capable of characterizing the dynamics of the system as
\begin{equation}
    \tilde{x}_{k+1}=\tilde{A}\tilde{x}_k
\end{equation}
The spatiotemporal coherent modes signifying the system dynamics are computed from the dominant eigendecomposition of the matrix $\tilde{A}$. 
\begin{equation}
    \tilde{A}W=W\Lambda 
\end{equation}
with $W$ and $\Lambda$ denoting the matrix of eigenvector and the diagonal matrix with the diagonal entries as eigenvalues respectively. 
These eigenvectors indicate the DMD modes which are of reduced dimensions. To reconstruct the subspace with original dimensions the exact DMD modes $\Phi$ are evaluated.
\begin{equation}
    \Phi =X_2V\Sigma ^{-1}W
\end{equation}
With the dominant eigenvalues and the exact DMD modes, the present and future states of the system are evaluated as
\begin{equation}
    x(t)=\sum_{k=1}^{r} \phi _kexp(\omega _kt)b_k=\Phi exp(\Omega t)b
    \label{solution}
\end{equation}
where $\omega _k=ln(\lambda _k)/\Delta t$ are continuous-time eigenvalues, $\Omega =dia(\omega )$ and $b=\Phi ^\dagger x_1$ represents the initial values of the DMD modes. 

\subsection{Prediction model analysis}
For the prediction model generalization, the identified approximate linear model (\ref{solution}) characterizing the innate dynamical behavior of the system is tested with the test data set which was not included with the training data set. 
The accuracy of the model is checked by calculating the root mean square errors (RMSE) for examining the prediction performance. The causal inference among the inputs is further analyzed with the statistics of the GCT test. The stronger temporal causal effect is analyzed with the test statistic distribution and the p-value. 

The PE condition will provide the lower bound on the length of the trajectory enabling the algorithm for capturing the essential modes of the system but will not ensure predictability. The predictability is only ensured when the temporal causal relationship is established with the help of GCT. %The Frobenius norm will become zero at the accurate identification of the approximate linear model.

\section{Results}\label{results}
The comprehensive insights into the proposed methodology are gained with the application to the coherency identification in multi-machine power systems (MMPS). 
The highly interconnected huge structure of MMPS integrated with renewable energy resources exhibits complex nonlinear dynamic behavior. During a disturbance, the groups of generators belonging to a particular geographical area in an MMPS tend to swing together with the same angular speed \cite{el2010identification} for maintaining relative power angles constant post transient which is known as the coherency property. 
During large disturbances, the power angles of the generator lose synchronism leading to the loss of the coherency property of the generators hence the generators may oscillate with a frequency of different groups. On that account for transient stability analysis and ensuing preventive control actions, the identification of coherency in the MMPS is necessary.

\subsection{DMD approach for Coherency Identification in MMPS}
The data-driven approach for identifying the coherency by capturing the transients in MMPS with coupled generators is demonstrated by considering an illustrative case study of a 22 bus 6 generator system.  The sufficiency of the data for the identification of the approximate linear model is assured with the condition of persistence of excitation and the predictability of the identified approximate linear model is tested with the Granger Causality. As the coherency detection of generators after a disturbance can be scrutinized during steady state conditions irrespective of underlying dynamics, the DMD methodology can be applied to capturing the coherency.

\subsubsection{Dataset}

\begin{figure}[ht!]
    \centering
    \includegraphics[width=\linewidth]{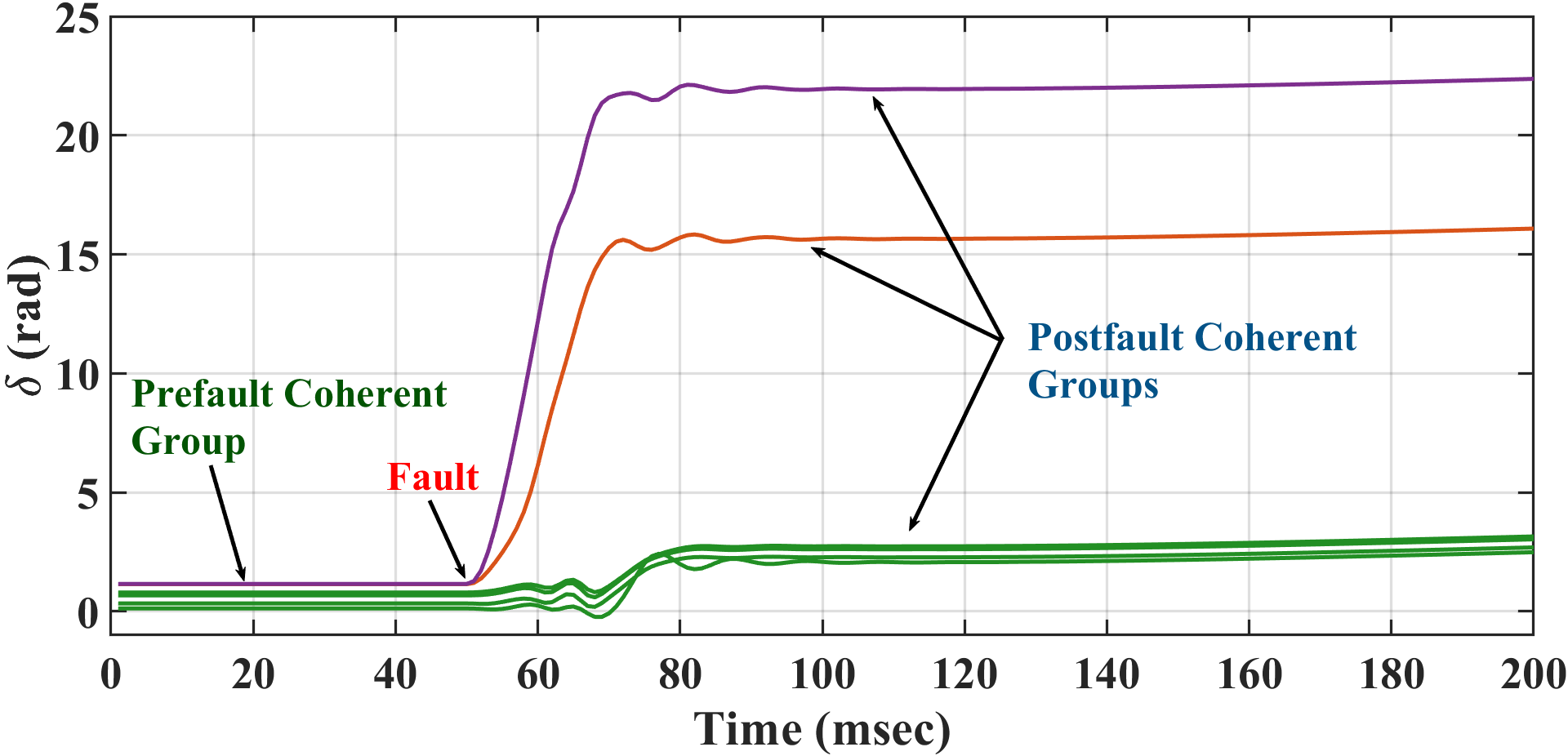}
    \caption{Rotor angle data of generators with pre-fault and post-fault coherency scenarios}
    \label{Data}
\end{figure}

The measurement data gathered from the Phasor Measurement Units (PMUs) installed in the power system network for determining the approximate linear model are obtained from \cite{wagh2009non}. Although coherency detection is a steady-state phenomenon, acquiring steady-state data during an event of large disturbance is difficult. Consequently, for capturing the overall dynamics of the system with the approximate linear model, the data up to transients are taken into account. The data for rotor angles of all generators to 200 milliseconds (msec) is utilized for identifying the underlying dynamical model of the system. Up to 50 msec, all six generators swing together in one coherent group. The real power supplied by each generator changed after 50 msec as a consequence of the line outage that occurred due to a fault or sudden variation in the load at 50 msec.
As described in Fig \ref{Data}, after 50 msec, one coherent group is separated into three clusters of different coherent groups, out of which one group is formed by four out of six generators, and the remaining two generators form individual coherent groups.

Before applying DMD, it is necessary to check the sufficiency of data for the identification of the linear model, which is achieved through checking two conditions viz PE condition and the GCT. The data should satisfy the PE condition and pass the GCT to provide an accurate estimate of the linear model approximating the dynamic behavior of the system with the capability of long-term predictability. The three cases demonstrating different scenarios with various hankel lags and the resulting prediction performance of the identified approximate linear model are discussed briefly henceforth.  

\subsubsection{Case 1: Data is not PE and GCT is not true}
\begin{figure}[ht!]
    \centering
    \includegraphics[width=\linewidth]{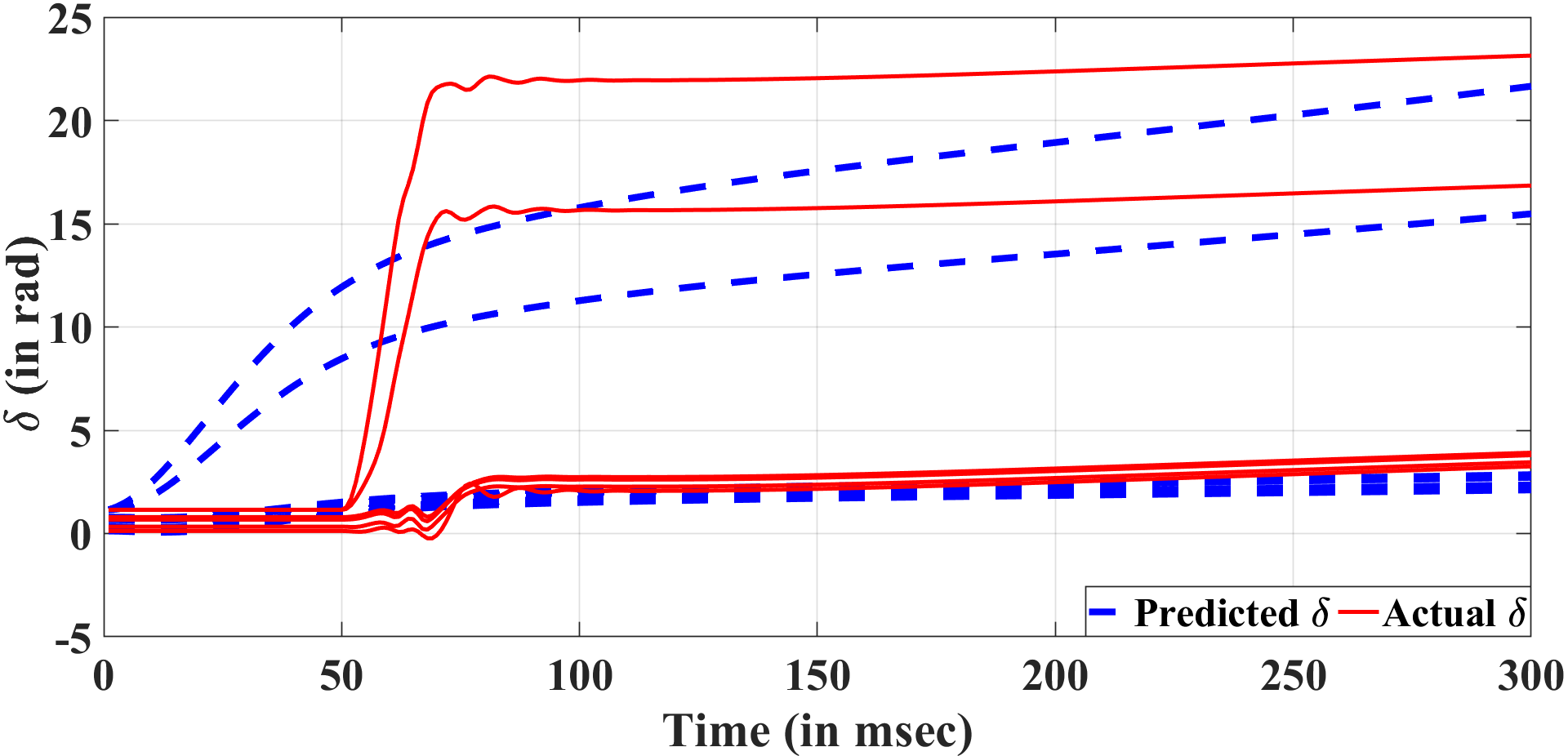}
    \caption{Prediction of rotor angle with DMD algorithm for coherency identification without using the hankel matrix }
    \label{L1}
\end{figure}

This case discusses the scenario where the data matrices for the DMD will contain one row with each measurement being a vector $x(k)\in \mathbb{R}^6$ whose element is the rotor angles corresponding to each of the six generators. For capturing the underlying dynamics of the system, the exact DMD modes are evaluated with the DMD algorithm. From Fig \ref{L1}, it is clear that the identified model fails to provide accurate tracking of the actual data due to data insufficiency. For the estimation of the best-fit system matrix $A$ through the DMD technique, it is necessary that the input data matrices should contain both the temporal and spatial measurement samples to evaluate the spatiotemporal coherent modes. The input matrices lack spatial measurements and contain temporal samples leading to a violation of the PE condition.
Besides, the cardinality of the measurement signal should be greater than the number of the dominant modes. In this illustration, the identified dominant modes are not sufficient to formulate the linear model. Furthermore, the GCT is invalid in such an event, leading to the inaccurate estimation of an approximate linear model depicting the system dynamics.

\subsubsection{Case 2: PE is satisfied but GCT is false}
\begin{figure}[ht!]
    \centering
    \includegraphics[width=\linewidth]{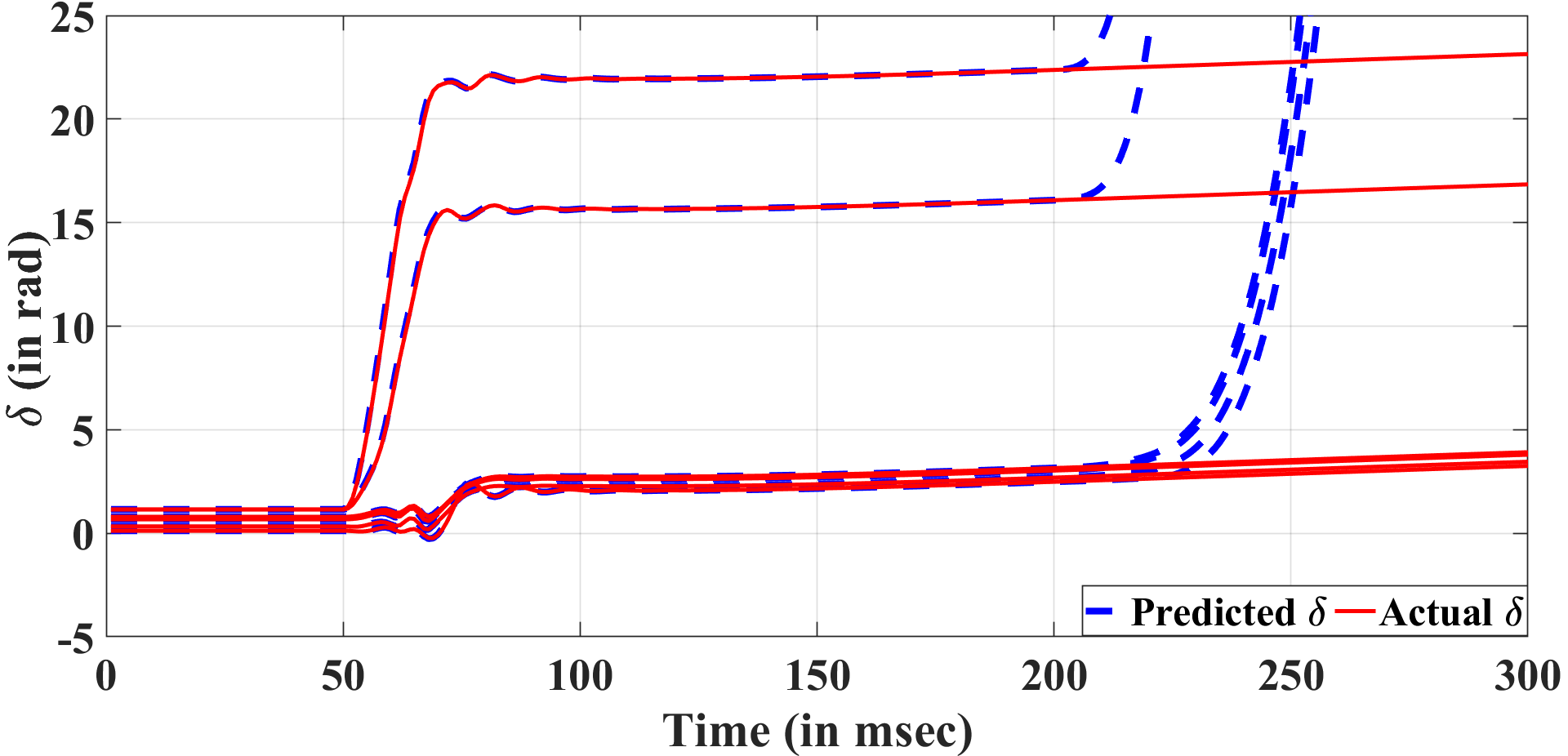}
    \caption{Rotor angle prediction for coherency identification using DMD with Hankel lag 50}
    \label{L50}
\end{figure}

\begin{figure}[ht!]
    \centering
    \includegraphics[width=\linewidth]{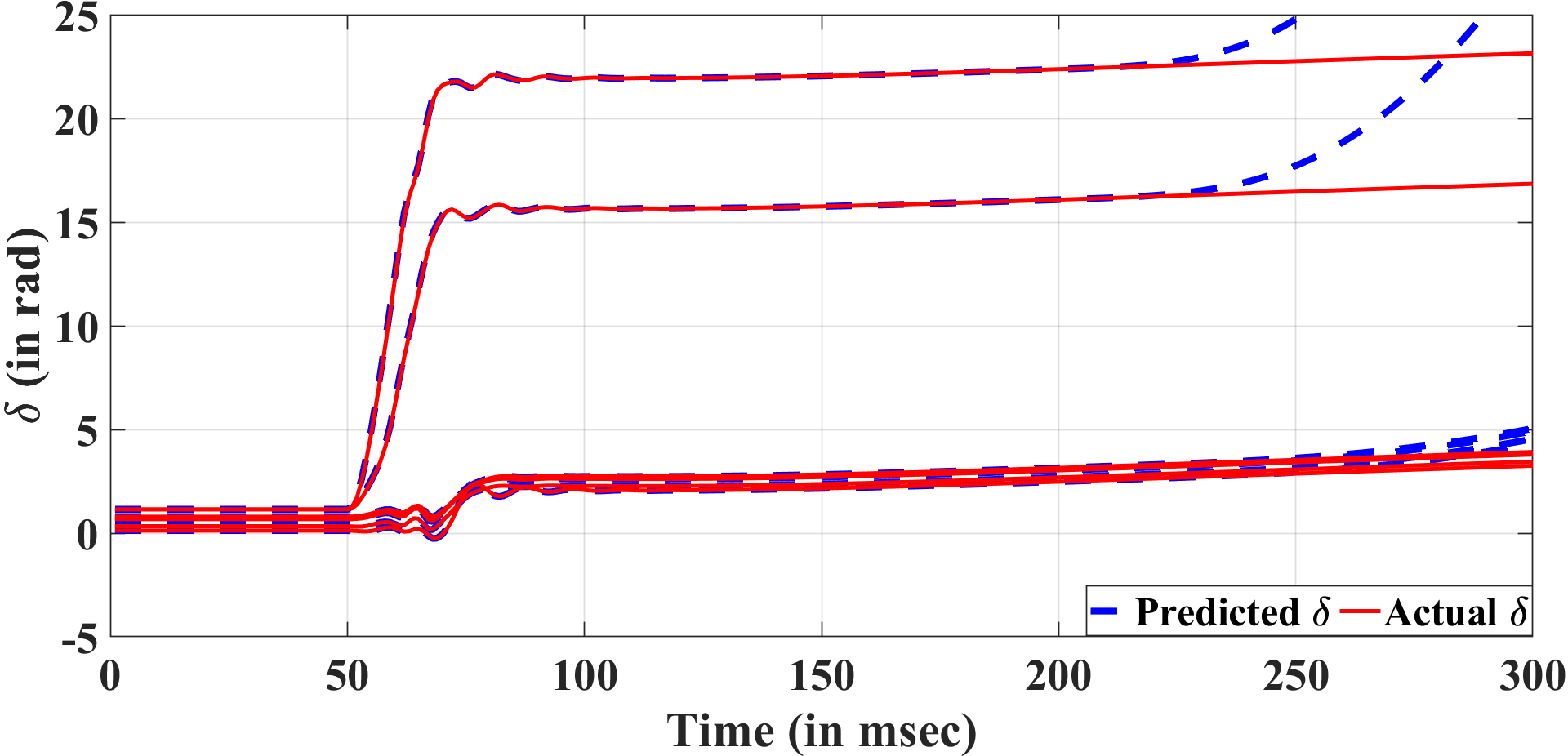}
    \caption{Rotor angle prediction for coherency identification using DMD with Hankel lag 51}
    \label{L51}
\end{figure}
Till 50 msec, there is hardly any change in the rotor angles depicting the steady state. The sudden change in rotor angle occurs after the fault at 50 msec; hence the model starts capturing the dynamics in the data at this moment. Therefore at this moment, when a sudden change in the measurements is observed, the order of PE should start. Hence this case discusses the scenario where the new state variables are introduced by increasing the order of the model with the help of the hankel matrix of lag 50 and lag 51, i.e., $(L=50)$ and $(L=51)$. In this case, the dimensions of input data matrices are increased such that there will be more temporal measurement samples than spatial samples to satisfy the PE condition. 
The model succeeded in capturing most of the spatiotemporal coherent modes, which could reconstruct the input dataset and track the actual rotor angle data to 200 msec. However, the estimated approximate linear model failed to provide prediction when tested for the further 100 msec data specifying that the estimated model failed at the predictability, which is also evident from the results of GCT. Furthermore, the test failed (i.e., the hypothesis testing returned logical 0 output specifying the absence of a causal relationship between the past and future measurements) when performed on the extended time series of past and future measurement samples, indicating that $X_2$ is not granger causing $X_1$ leading to the inaccurate prediction.  

\subsubsection{Case 3: PE is fulfilled and GCT is true}

\begin{figure}[ht!]
    \centering
    \includegraphics[width=\linewidth]{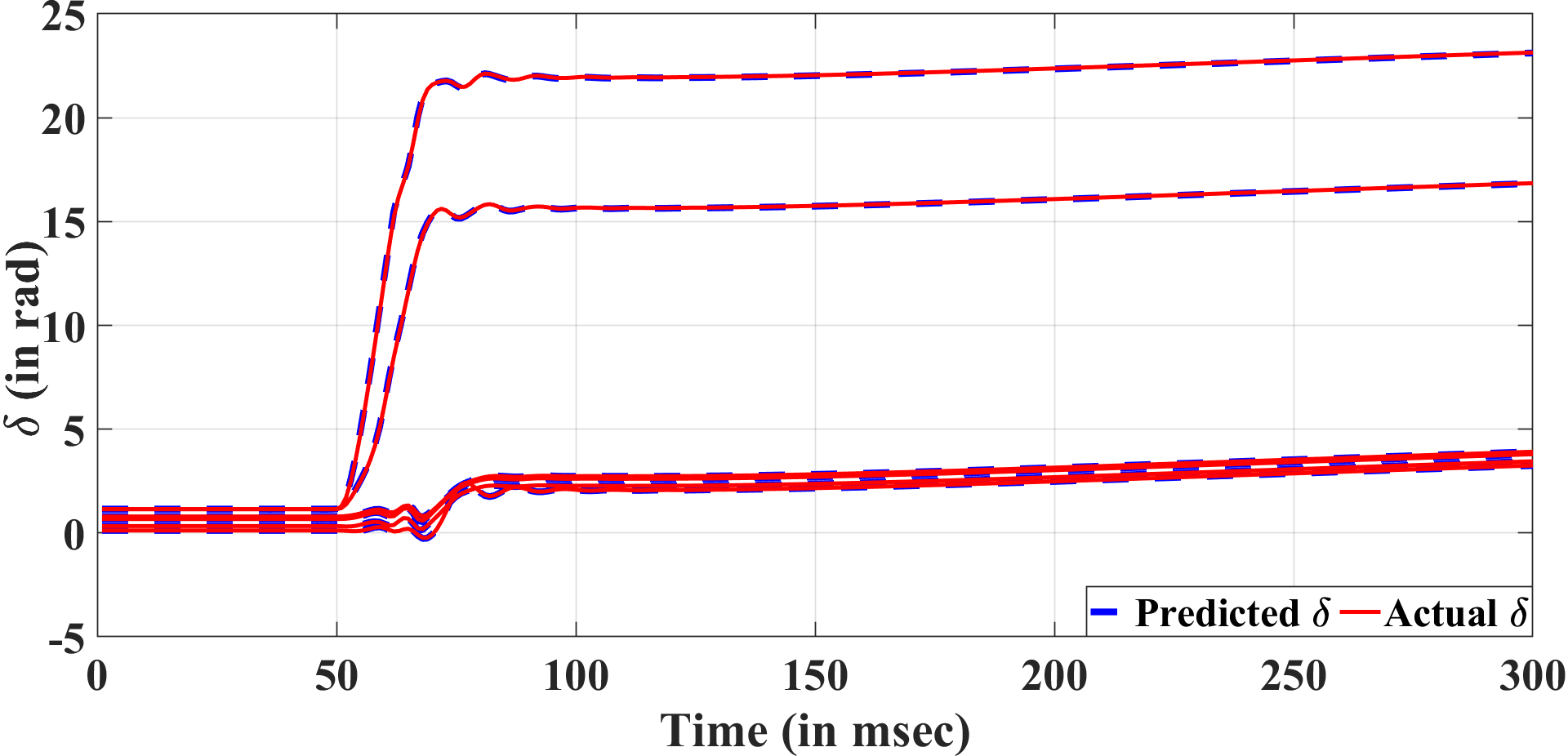}
    \caption{Prediction of rotor angle for coherency identification using DMD with Hankel lag 51}
    \label{L52}
\end{figure}

 In a nutshell, this case discusses the scenario when the PE condition is satisfied, and the GCT is also passed. The dimensions of the input matrices are further varied by increasing the hankel lag to 52 $(L=52)$ satisfying the PE condition. Also, when the GCT is performed on these extended time series with the virtually added state variables, the test returns logical one output, or the GCT is true, implying that there is a causal relationship between the past and future measurements (i.e., $X_2 $ is granger causing $X_1$). As a result of both the conditions PE and GCT being satisfied, the criterion for the data validation is fulfilled, and the accurate linear model with predictability is estimated, which is evident from Fig. \ref{L52}, where the rotor angle prediction accurately tracks the actual rotor angle measurements for both the training (200 msec) and testing data points (100 msec).

\subsection{Eigen Value Analysis}
% \begin{figure*}[ht!]
%     \centering
%     \includegraphics[width=0.95\linewidth]{Figures/eigen_plots-3.pdf}
%     \caption{Eigen value analysis corresponding to various hankel lags}
%     \label{EA}
% \end{figure*}
\begin{figure}[ht!]
    \centering
    \includegraphics[width=\linewidth]{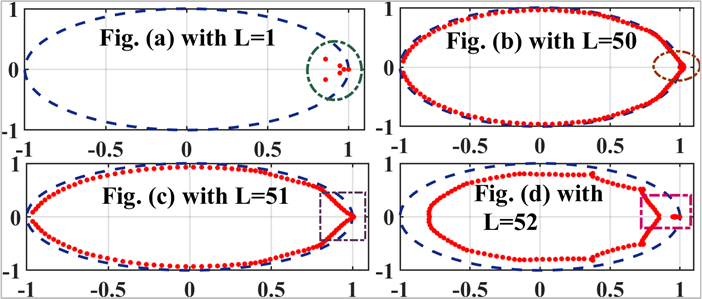}
    \caption{Eigen value analysis corresponding to various hankel lags}
    \label{EA}
\end{figure}
\begin{figure}[ht!]
    \centering
    \includegraphics[width=\linewidth]{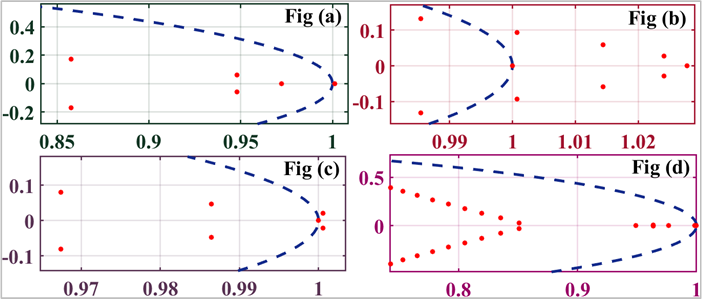}
    \caption{Zoomed version of eigenvalues corresponding to various hankel lags}
    \label{EA_zoomed}
\end{figure}
%  The SVD requires the inverse of the diagonal matrix of singular values $\Sigma^{-1}$ which poses a numerical problem, as these small singular values may be inaccurate. applying $\Sigma^{-1}$ could overamplify components of $X_1$, in particular, the less important ones. 

As observed from Fig. \ref{EA} (a), the eigenvalues extracted from the spatiotemporal decomposition of data without any Hankel matrix representation are not enough to build the approximate linear model for the predictability. As the subspace spanned by the measurements is increased with the Hankel representation, the number of identified eigenvalues starts increasing. At the hankel lag 50 (Fig. \ref{EA_zoomed} (b)), the evaluated 50 eigenvalues are capturing the system dynamics and hence providing the reconstruction of the training data but some of the eigenvalues lie outside the unit circle leading to some instability in the estimated model resulting into inaccurate predictions. Further increasing the hankel lag to 51 (Fig. \ref{EA_zoomed} (c)) some of the eigenvalues still lie outside the unit circle. When the hankel lag is increased to 52 (Fig. (\ref{EA_zoomed} (d)) all the dominant eigenvalues characterizing the dynamic behavior of the system lie inside the unit circle ensuring the estimation of the accurate approximate linear model with predictability.

\subsection{Granger Causality Test Analysis}
\begin{table}[ht!]
\caption{Granger Causality Test Results with varying hankel lag}
\label{GCT results}
\begin{center}
\begin{tabular}{|c|c|c|c|c|}
\hline
\textbf{Case} & \textbf{L}  & \textbf{Test} & \textbf{p value} & \textbf{Test Statistic} \\ \hline
1    & 50 & 0    & 0.4390  & 0.6            \\ \hline
2    & 51 & 0    & 0.1690  & 1.89           \\ \hline
3    & 52 & 1    & 0.0243  & 5.08           \\ \hline
4    & 54 & 1    & $9.42\times 10^{-4}$    & 10.9           \\ \hline
5    & 58 & 1    & $1.86\times 10^{-4}$    & 14             \\ \hline
6    & 60 & 1    & $5.36\times 10^{-8}$    & 29.6           \\ \hline
\end{tabular}
\end{center}
\end{table}

\begin{figure}[ht!]
    \centering
    \includegraphics[width=\linewidth]{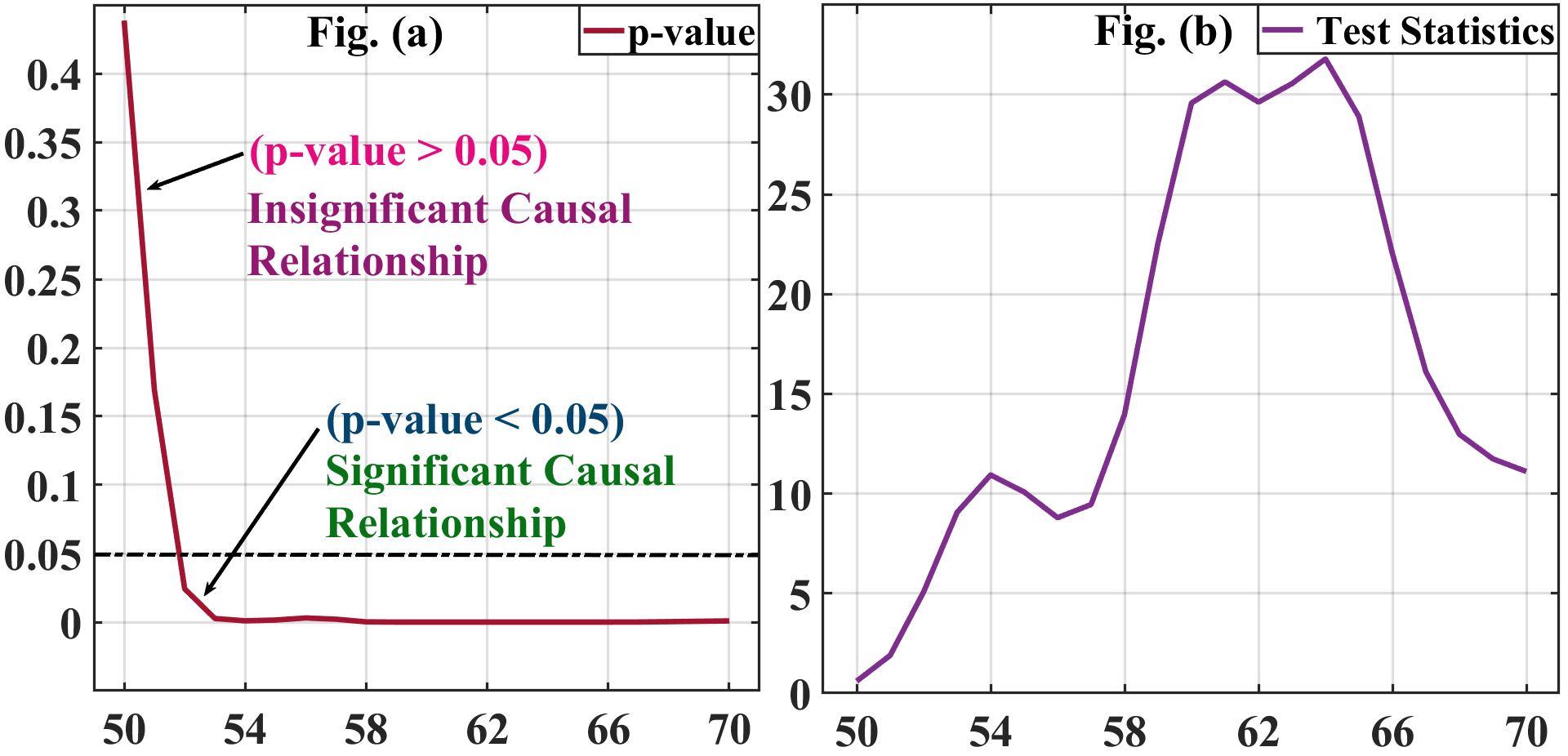}
    \caption{GCT analysis: p-value and test statistics}
    \label{pval}
\end{figure}

The p-value quantifies the probability of procuring the observed results with the assumption that the null hypothesis is true. The p-value in hypothesis testing signifies the rejection of a null hypothesis so long as the evidence in favor of the rejection is available. A smaller p-value implies the availability of stronger evidence in the favour of the alternative hypothesis. Generally, a p-value lesser than 0.05 indicates a statistically significant relationship. In GCT the null hypothesis represents a noncausal relationship whereas the alternative hypothesis defines the existence of Granger causality. Table \ref{GCT results} summarises the results of the hypothesis testing implemented for examining the Granger causality for predictability in DMD with various hankel lags. It is observed that the larger p-value for $L=50$ and $L=51$ indicates the decision in favor of the null hypothesis signifying that there is no causal relationship resulting in inaccurate predictions. As the lag of Hankel matrices has increased the p-value for $L=52$ decreased to 0.0243 which is lesser than 0.05 indicating stronger evidence in favor of rejecting the null hypothesis of non-causality. With the repeated experiments including various hankel lags $L=54$, $L=58$ and $L=60$ the p-value observed is lesser than 0.05 and assured the statistically significant existence of Granger causality indicating the accurate estimated data-driven models with predictability. The p-value evaluated from the repeated experiments  with increasing hankel lags from 50 to 70 is plotted in Fig. \ref{pval} (a) which summarises the results discussed above.

The test statistic is a random variable used for evaluating the p-value by examining the agreement between the samples and the null hypothesis. Test statistic provides the information concerning data in accordance with the decision of null hypothesis rejection. The sampling distribution of test statistics is illustrated in Fig. \ref{pval} (b). It is observed that for the scenarios concerning the strong evidence in favor of the alternative hypothesis, the magnitude of test statistics shows too large or too small values relevant to the alternative hypothesis causing a sudden decrease in the p-value implying the rejection of the null hypothesis.

\subsection{Condition Number associated with various hankel lags}
% \begin{table}[ht!]
% \caption{Condition numbers corresponding to various hankel lags}
% \begin{tabular}{|c|c|c|c|c|}
% \hline
% \textbf{Case} & 1     & 2     & 3     & 4     \\ \hline
% \textbf{L}    & 50    & 52    & 54    & 60    \\ \hline
% \textbf{k(A)} & $5.599 \times 10^9$ & $7.441 \times 10^6$ & $1.888 \times 10^6$ & $1.818 \times 10^6$ \\ \hline
% \textbf{k(w)} & $7.661 \times 10^9$ & $3.333 \times 10^4$ & $1.314 \times 10^3$ & $2.702 \times 10^3$ \\ \hline
% \end{tabular}
% \end{table}

\begin{table}[ht!]
 \caption{Condition numbers corresponding to various hankel lags}
\begin{tabular}{|c|c|c|c|}
\hline
\textbf{Case} & \textbf{L} & \textbf{k(A)}  \\ \hline
1             & 50         & $5.599 \times 10^9$            \\ \hline
2             & 52         &  $7.441 \times 10^6$               \\ \hline
3             & 54         & $1.888 \times 10^6$             \\ \hline
4             & 60         & $1.818 \times 10^6$              \\ \hline
\end{tabular}
\label{Cond Number}
\end{table}
% \begin{table}[ht!]
%  \caption{Condition numbers corresponding to various hankel lags}
% \begin{tabular}{|c|c|c|c|}
% \hline
% \textbf{Case} & \textbf{L} & \textbf{k(A)} & \textbf{k(w)} \\ \hline
% 1             & 50         & $5.599 \times 10^9$         & $7.661 \times 10^9$         \\ \hline
% 2             & 52         &  $7.441 \times 10^6$         &  $3.333 \times 10^4$      \\ \hline
% 3             & 54         & $1.888 \times 10^6$        & $1.314 \times 10^3$         \\ \hline
% 4             & 60         & $1.818 \times 10^6$       & $2.702 \times 10^3$         \\ \hline
% \end{tabular}
% \end{table}

The condition number $k(A)$ quantifies the singularity existing among the matrix $A$. When the matrix $A$ is singular the condition number becomes infinite. The condition number quantifies the ratio of maximum relative stretching to the maximum relative shrinking that the matrix does to any non-zero vectors. 
The condition number is calculated as 
\begin{equation}
    k(A)=\left \| A \right \|\left \| A \right \|^{-1}=\left ( \underset{x\neq 0}{max} \frac{\left \| Ax \right \|}{\left \| x \right \|}\right ).\left ( \underset{x\neq 0}{min} \frac{\left \| Ax \right \|}{\left \| x \right \|}\right )^{-1}
\end{equation}
As observed from Table \ref{Cond Number}, the condition numbers for lag $L=52$ were large indicating the presence of singularities within the data. With the increasing hankel lags, the condition numbers for lag $L=52$  onward drop indicating the enriched representation of the data.
\subsection{Error Analysis}
\begin{table}[ht!]
\caption{Error Analysis with varying hankel lag}
\label{error analysis}
\begin{center}

\begin{tabular}{|cc|c|c|c|c|}
\hline
\multicolumn{2}{|c|}{\textbf{Case}}                               & 1       & 2       & 3      & 4      \\ \hline
\multicolumn{2}{|c|}{\textbf{L}}                                  & 50      & 52      & 54     & 60     \\ \hline
\multicolumn{1}{|c|}{\multirow{6}{*}{\textbf{RMSE}}} & \textbf{$\delta_1$} &  51.3651&  0.0097& 0.0123 & 0.0203 \\ \cline{2-6} 
\multicolumn{1}{|c|}{}                               & \textbf{$\delta_2$} & 599 & 0.0279& 0.0597 & 0.1242 \\ \cline{2-6} 
\multicolumn{1}{|c|}{}                               & \textbf{$\delta_3$} &  917  & 0.0402    & 0.083  & 0.1733 \\ \cline{2-6} 
\multicolumn{1}{|c|}{}                               & \textbf{$\delta_4$} &  56.40 &  0.0074  & 0.0143 & 0.0249 \\ \cline{2-6} 
\multicolumn{1}{|c|}{}                               & \textbf{$\delta_5$} &53.30  & 0.0082  & 0.0134 & 0.0219 \\ \cline{2-6} 
\multicolumn{1}{|c|}{}                               & \textbf{$\delta_6$} & 55.74 &  0.0087  & 0.0147 & 0.0245 \\ \hline
\end{tabular}
\end{center}
\end{table}
For analyzing the prediction performance of the identified approximate linear model the root mean square errors (RMSE) are evaluated as
\begin{equation}
    RMSE=\sqrt{\frac{1}{n} \sum_{i=1}^{n}(\hat{\delta_i}-\delta_i)^2}
\end{equation}
where $\hat{\delta}$ represents the predicted value of the rotor angle with DMD and $\delta$ represents the actual value of the rotor angle.
From Table \ref{error analysis}, it is evident that when only PE was satisfied and GCT was false, the identified model was able to reconstruct the training data but was unable to provide further prediction which is reflected from the larger values of RMSE corresponding to the hankel lag 50. With the increasing hankel lags, both the conditions PE and GCT are satisfied, leading to the accurate prediction reflected from the smaller RMSE values corresponding to hankel lags 52,54 and 60.

\section{Conclusion}\label{conclusion}
For capturing the dominant spatiotemporal coherent modes, the dataset should contain an adequate number of rows and columns, ensuring the sufficient order of the model. The dimensions of the subspace containing the underlying system modes spanned by the measurement data sequences were increased with the Hankel matrix representation adding the virtual state variables to the data. The lower bound on the length of the section of trajectory reproducing the actual system dynamics was found through the PE condition. The experimental case studies observed that satisfying the PE condition failed to guarantee predictability. The predictability of the identified model satisfying the PE condition was ensured with the causal relationship among the data established through the GCT. The prediction model's performance was interpreted with the eigenvalue analysis (ensuring that for accurate prediction, the eigenvalues lie within the unit circle), the GCT analysis (ensuring the stronger causal relationship with the lower p-values), and the error analysis. Ensuring the predictability of the nonlinear system with the GCT is a challenging task; hence, applying the Koopman operator for capturing the nonlinear dynamics through the linear evaluation of the state space functions is proposed in the future scope. Further, DMD could be employed for approximating infinite dimensional linear space, and the GCT could be employed to test the predictability. 
% \section*{Acknowledgment}
% The authors of the paper would like to acknowledge the support of Dr. N. M. Singh, Adjunct Prof., Control and Decision Research Center (CDRC), EED, VJTI, Mumbai, India, for providing guidance and rigorous technical discussion.
% The authors also would like to extend gratitude to the Dr. S. R. Wagh for providing support and  help in the  preparation of manuscript.

\bibliographystyle{IEEEtran}
\bibliography{citation}

% \appendix

\end{document}